\documentclass{aa}      
\usepackage{graphicx}

\def\m{\phantom -}
\def\:{\phantom :}
\def\1{\phantom 1}
\def\m100{mag 100d$^{-1}$}

\begin{document}
\thesaurus{6         
        (19.92.1  
         19.94.1  
                   }         

\title{THE ASIAGO SUPERNOVA CATALOGUE - 10 YEARS AFTER}
\titlerunning{The Asiago Supernova Catalogue 1999}
\author{R.Barbon\inst{1}, V.Buond\'\i\inst{1},  E.Cappellaro\inst{2}, 
M.Turatto\inst{2}}
\authorrunning{Barbon et al.}
\offprints{R. Barbon}

\institute{Osservatorio Astrofisico dell'Universit\`a di Padova, 
I-36012 Asiago, Italy \and 
Osservatorio Astronomico di Padova, vicolo dell'Osservatorio 5, 35122
Padova, Italy}
\date{Received ................; accepted ................}

\maketitle

\begin{abstract}
Ten years after the publication of the previous release, we present a
new edition of the Asiago Supernova Catalogue updated to December 31,
1998 and containing data for 1447 supernovae and their parent
galaxies\footnote{Tables 1 and 2 are only available in electronic form
at the CDS via anonymous ftp to cdsarc.u-strasbg.fr (130.79.128.5) or via
http://cdsweb.u-strasbg.fr/Abstract.html}.  In addition to list the data for a large number of new SNe,
we made an effort to search the literature for new information on past
SNe as well. We also tried to update and homogenize the data for the parent
galaxies.  To allow a global view of the Catalogue, a few descriptive
figures and a summary table is reported.  The present Catalogue is
intended as a large and modern database for statistical
studies on the supernova phenomenon.
\end{abstract}


\section{Introduction}

The interest of the scientific community on supernovae (SNe) has
enormously increased in the recent years for several reasons. The
advances in the understanding of the SN phenomena obtained with the
intensive study of nearby SNe, first of all SN~1987A, have raised new
more fundamental questions with regard to progenitor evolution,
explosion mechanism and nucleosynthesis.  In addition, the calibration
of the absolute magnitudes of a few SNIa obtained using the Cepheid
variables found in their parent galaxies (Saha et al., 1999, and references therein), and the discovery of
empirical relations between the absolute magnitudes at maximum and the
shape of the light curves of SNIa (Phillips 1993, Riess et al. 1996)
have renewed the interest for the utilization of SNIa as distance
indicators up to cosmological distances.  Other exciting advances are
expected for the association of some SNe with the mysterious
GRBs. Such wide interest has triggered new, deep SN searches which, in
a few years have doubled the number of SN discoveries.

The history of the Asiago SN catalogue began in 1984 with the
publication of data for 568 objects (Barbon et al. 1984). This was
compiled starting from the Palomar Supernova Master List which since
1958 from time to time appeared in the literature (Zwicky 1958 and
1965, Kowal and Sargent 1971, Sargent et al. 1974). During the same
period two other SN listing have been published, by Karpowicz and
Rudnicki (1968) and by Flin et al. (1979), the latter giving also the
complete bibliography for each object. 
The 1984 Asiago SN Catalogue was superseded by a new edition in 1989
(Barbon et al. 1989, [ASC89]) which listed information for the 661
supernovae discovered up to December 31, 1988.

More recently, van den Bergh (1994) published a list containing the
203 supernovae discovered between January 1, 1989 and April 3, 1994
and a {\em Catalogue of extragalactic Supernovae}, complete up to
1993, was published in volume V of the {\em General Catalogue
of Variable Stars} (Samus 1995). 

In the last few years we made available through the WEB at the URL
{\em athena.pd.astro.it/$\sim$supern/} a running SN list which was widely
utilized through the literature. Other supernova listings are
available electronically, e.g. the list at the CBAT ({\em
www.harvard.edu/\-iau/\-lists/\-Super\-novae.html}), and that at
Sternberg Astronomical Institute ({\em
www.sai.msu.su/\-cgi-bin/\-wdb-p95/\-sn/\-sncat/\-form}).

The many requests for a new, reliable edition give the motivation for
the preparation of the present paper.

\section{The Catalogue}

The new edition of the Asiago SN Catalogue lists data for 1447 SNe and
for their parent galaxies discovered up to 31 December, 1998.  For the
galaxy data we made large use of the {\em Third Reference Catalogue of
Bright Galaxies} by de Vaucouleurs et al. (1991, [RC3]) and of the
LEDA \footnote{LEDA Lyon-Meudon Extragalactic Database;
www-obs.univ-lyon1.fr} and NED\footnote{The NASA/IPAC Extragalactic
Database (NED) is operated by the Jet Propulsion Laboratory,
California Institute of Technology, under contract with the National
Aeronautics and Space Administration.} databases.
 
The format of the new edition follows that of ASC89 with some
improvements. In particular, we have now included:

\begin{enumerate}
\item accurate supernova positions.
\item position angles of the major axes. 
\item morphological type code of the parent galaxies.
\end{enumerate}

Accurate SN positions are mostly useful to compare observations at
different wavelengths, from X-ray to radio, which in the recent years
had a large impact in SN research. Note that in the present Catalogue
all coordinates are given at the 2000.0 epoch.
Major axis position angles were introduced to study the position of
the SNe within the galaxies and numerical morphological type code to
facilitate the derivation of descriptive statistics.

Instead, we choose to drop the information on the parent galaxy
luminosity classes because this information is available only for a small fraction (less than 20\%) of the objects. The
galaxy integrated luminosity can be computed
from the apparent magnitudes and distances of the galaxies.

To facilitate the consultation of the Catalogue, we present it with two
different sortings: in Table 1 the list is arranged chronologically
according to the date of SN discovery while in Table 2 the same data
are listed in order of Right Ascension.

In the Tables, the content of the different columns is as follows:

\begin{description}

\item[\bf{ 1:}] supernova designation. The symbols ``?'' denote a not
confirmed SN and ``*'' the occurrence of multiple SN discoveries in the
same galaxy.

\item[\bf{ 2:}] 
parent galaxy identification. In case a galaxy has different
identifications, we adopted the following priority: NGC, IC, MCG (M),
UGC, ESO (E), PGC, Leda, others. In some cases specific names are
reported, e.g. LMC. Anonymous galaxies are listed with the letter A
followed by the coordinates. In a few cases, where the association
with a definite parent galaxy was not possible, we have filled the
field with INTERGALACTIC.

\item[\bf{3-4:}] equatorial coordinates of the parent galaxy at the 2000.0
epoch.

\item[\bf{5-6:}] equatorial coordinates of the supernova at the 2000.0 epoch.

\item[\bf{7:}] morphological type of the parent galaxy.

\item[\bf{8:}] morphological type code for the parent galaxy (coding as in 
RC3)

\item[\bf{9:}] only for disk-like system, inclination of the polar axis 
 with respect to the line of sight in degrees ($0$ for face on
 systems).

\item[\bf{10:}] position angle of the major axis of the parent galaxy 
(North Eastwards) in degrees.

\item[\bf{11:}] heliocentric radial velocity of the parent in km~s$^{-1}$,
but for objects with redshift $z \geq 0.1$ where the $z$ value has been
listed.

\item[\bf{12:}] integrated B magnitude of the parent, mostly from the RC3
or LEDA. In few cases only photographic magnitudes (prefixed by ``p'')
are available.

\item[\bf{ 13:}] decimal logarithm  of the apparent isophotal diameter, in 0.1
arcmin units.

\item[\bf{14-15:}] SN offset from the galaxy nucleus in arcsec, in 
the E/W and N/S direction respectively.

\item[\bf{ 16:}] if available, supernova magnitude at maximum 
(photometric band indicated); otherwise discovery magnitude (labelled
by ``*'').  A magnitude
without band means that the observation has not been made in a
standard photometric system (e.g. those reported in the discovery
announcement as photographic, blue plate, red plate, CCD without
filter, and so on ).

\item[\bf{17:}] supernova type, mostly from spectroscopy. In a few 
cases, marked by ``*'', types have been inferred from the light
curve.

\item[\bf{18:}] if known, epoch of maximum, otherwise ``*'' marks date of 
discovery.

\item[\bf{19:}] name(s) of discoverer(s). For organized search teams
the acronyms are given.
\end{description}

\begin{table*}
\caption{Sample page of the Asiago Supernova Catalogue}
\resizebox{\hsize}{!}{\includegraphics*[angle=180.,bbllx=30,bblly=30,
bburx=550,bbury=800]{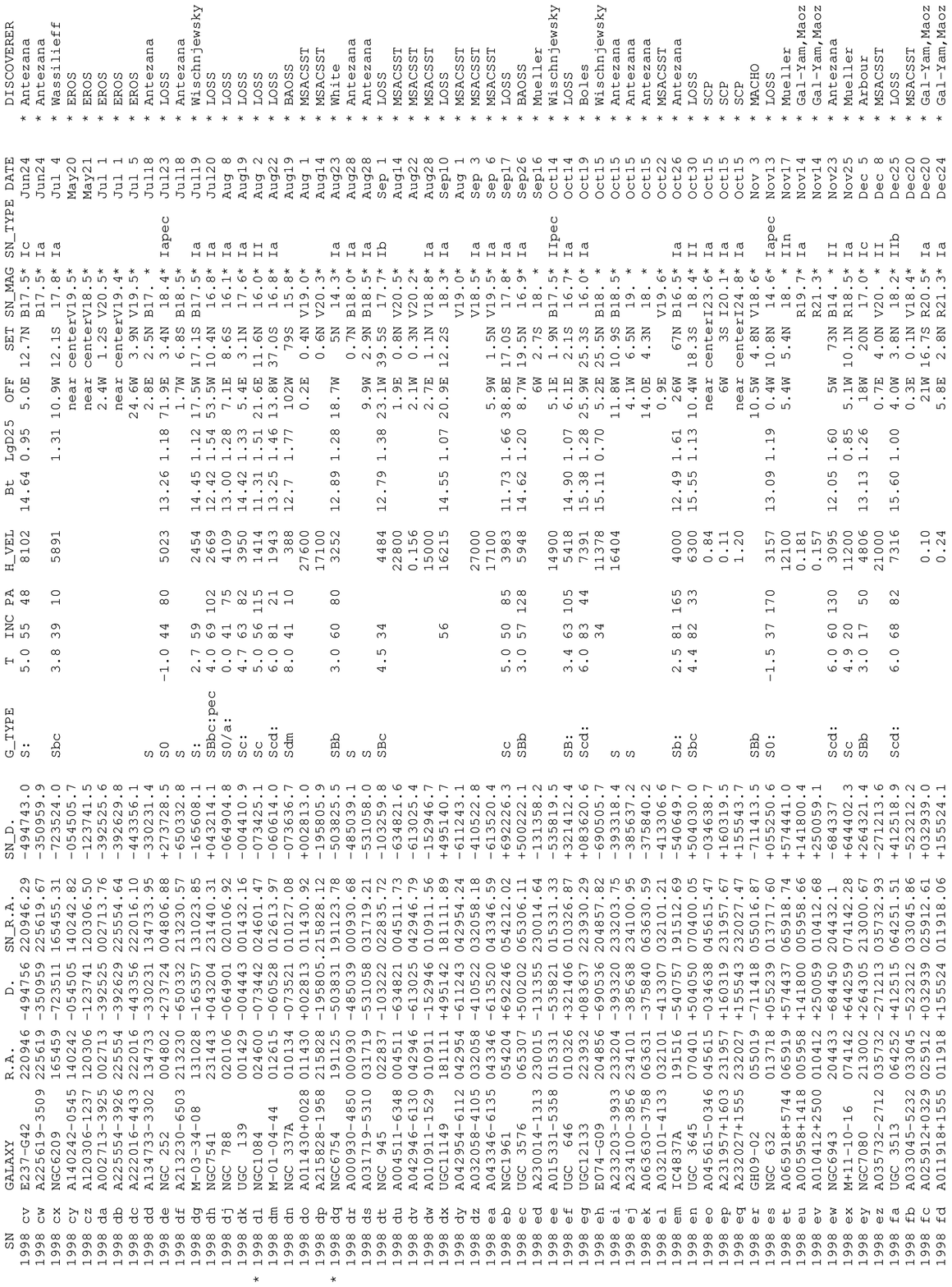}}
\end{table*}

\section{Remarks}
A major effort has been devoted in searching the literature for
accurate magnitudes, epochs of maximum and for assigning supernova
types. For this latter task we are indebted to D. Branch (private
communication) who provided us with a list of revised supernova types
for many SNe. For the supernovae discovered in the periods from 1989
to 1992 and from 1992 to 1998 we have cross checked our data with
those of van den Bergh (1994) and the electronic list supplied by the
CBAT,
respectively. Discrepancies, in both cases, have been solved by
looking at the original literature, mostly consisting of IAU
circulars.

The galaxy coordinates are given with various degrees of accuracy
depending on the accuracy of the original Catalogue. For many anonymous
galaxies, in particular for the parent galaxies of high-z SNe, the
approximate galaxy position is derived from the supernova
coordinates. In some cases, discrepancies may arise between the quoted
off-sets of the supernova from the galaxy nucleus and the same data
derived from the SN-parent relative coordinates (see
e.g. SN1965C). This happens especially for parent galaxies with ill
defined nuclei. Finally, a few supernovae have been discovered and
observed only spectroscopically (e.g SN1995bb) and therefore they lack a photometric magnitude.

In Table 3 we list the 37 SNe (excluding SN 1998ab,
added in the proofs) which, though announced after the publication of
ASC89, where discovered on old plates obtained before December 31, 1988.
 
Instead, Table 4 lists the 20 objects which, from the beginning of the
searches up to the present time, turned out not to be
supernovae. Among these latter objects, SN1950E and 1956C were still
included in ASC89. SN~1987G has been deleted because it turned out to be 
the same object as SN~1987D.

\setcounter{table}{2}
\begin{table}
\tabcolsep 0.10truecm
\caption{Supernovae not included in the ASC89.}
\begin{tabular}{lll}
\hline
 1945B~~~~~~~~&   1975U?~~~~~~~~& 1983ab~~~~~~~~\\
 1950N?&   1975V?&  1985S  \\
 1950O?&   1976O?&  1985T? \\
 1951J?&   1976P?&  1986P? \\
 1953J?&   1976Q?&  1987Q? \\
 1953K?&   1977I?&  1987R? \\
 1953L?&   1978J &  1987S  \\
 1954ac?&  1978K &   1988ac\\
 1954ad?&  1978L?&  1988ad?\\
 1955Q?&   1980Q &  1988ae\\ 
 1955R?&   1982Z?&  1988af\\ 
 1955S?&   1982aa&\  1988ag\\
 1962R?&   \\
\hline
\end{tabular}
\end{table}

\begin{table}
\tabcolsep 0.10truecm
\caption{Objects which turned out not to be real supernovae}
\begin{tabular}{lll}
\hline
 1950E~~~~~~~~ &  1986F~~~~~~~~ & 1991W~~~~~~~~ \\
 1956C &  1986H & 1991ap\\
 1967F &  1987E & 1992W  \\
 1973G &  1987H & 1993U \\
 1893H &  1988X & 1993V \\
 1985J &  1990C & 1998di \\
 1986D &  1992X\\
\hline
\end{tabular}
\end{table}

\section{Basic statistics} 

\begin{figure}
\resizebox{\hsize}{!}{\includegraphics*[angle=-90.]{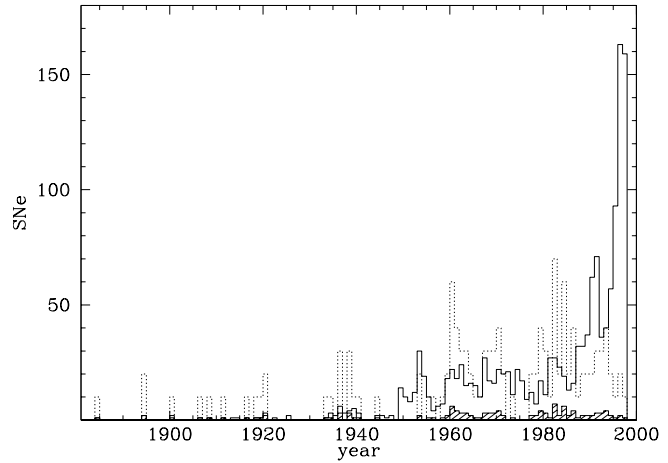}}
\caption{Histogram of the number of SNe discovered per year. The shaded area
refers to SNe with magnitude at maximum (or at discovery) brighter than
14 which, enlarged by a factor 10, are also shown as dotted lines.
}\label{disco}
\end{figure}

In Fig.~\ref{disco} the productivity of world wide supernova searches through
the years is shown. The enormous improvement of the last few years
stands out clearly.  The dashed area refers to SNe brighter than
14 mag. It must be noted that, despite the renewed effort in
SN search, the rate of discovery of bright SNe is not increasing.

\begin{figure}
\resizebox{\hsize}{!}{\includegraphics*[angle=-90.]{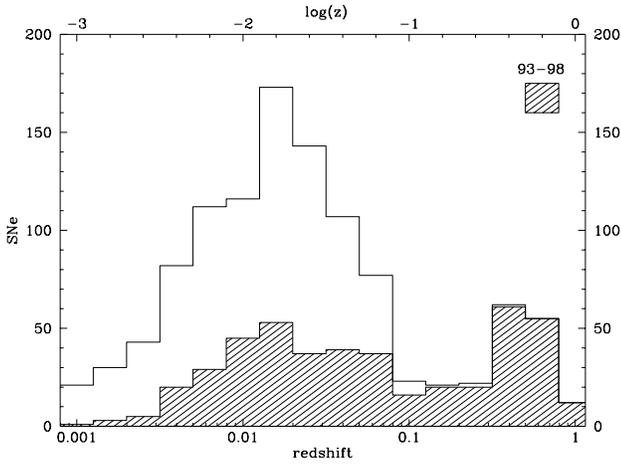}}
\caption{Distribution of SNe with the redshift of the host galaxy. The shaded area is
relative to the SNe discovered in the last 6 years.}\label{redsh}
\end{figure}

In Fig.~\ref{redsh} we plot the distribution of redshifts and
highlight (shaded area) the SNe discovered in the last 6 years.  It
appears that almost all SNe at redshift $z>0.1$ have been discovered
recently and that the recent high-z surveys favour discoveries in the
range $0.3 < z < 0.8$.

\begin{figure}
\resizebox{\hsize}{!}{\includegraphics*{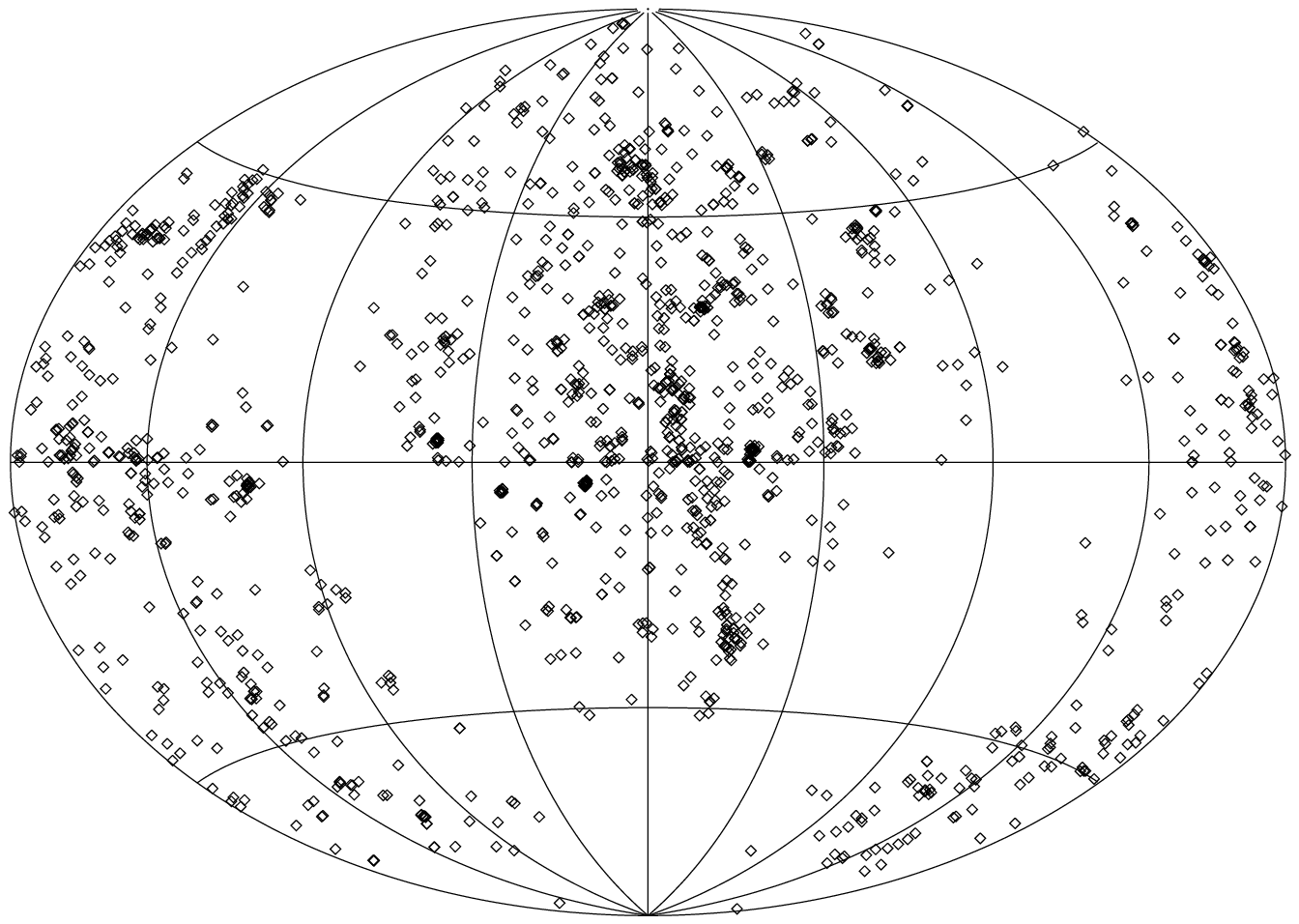}}
\resizebox{\hsize}{!}{\includegraphics*{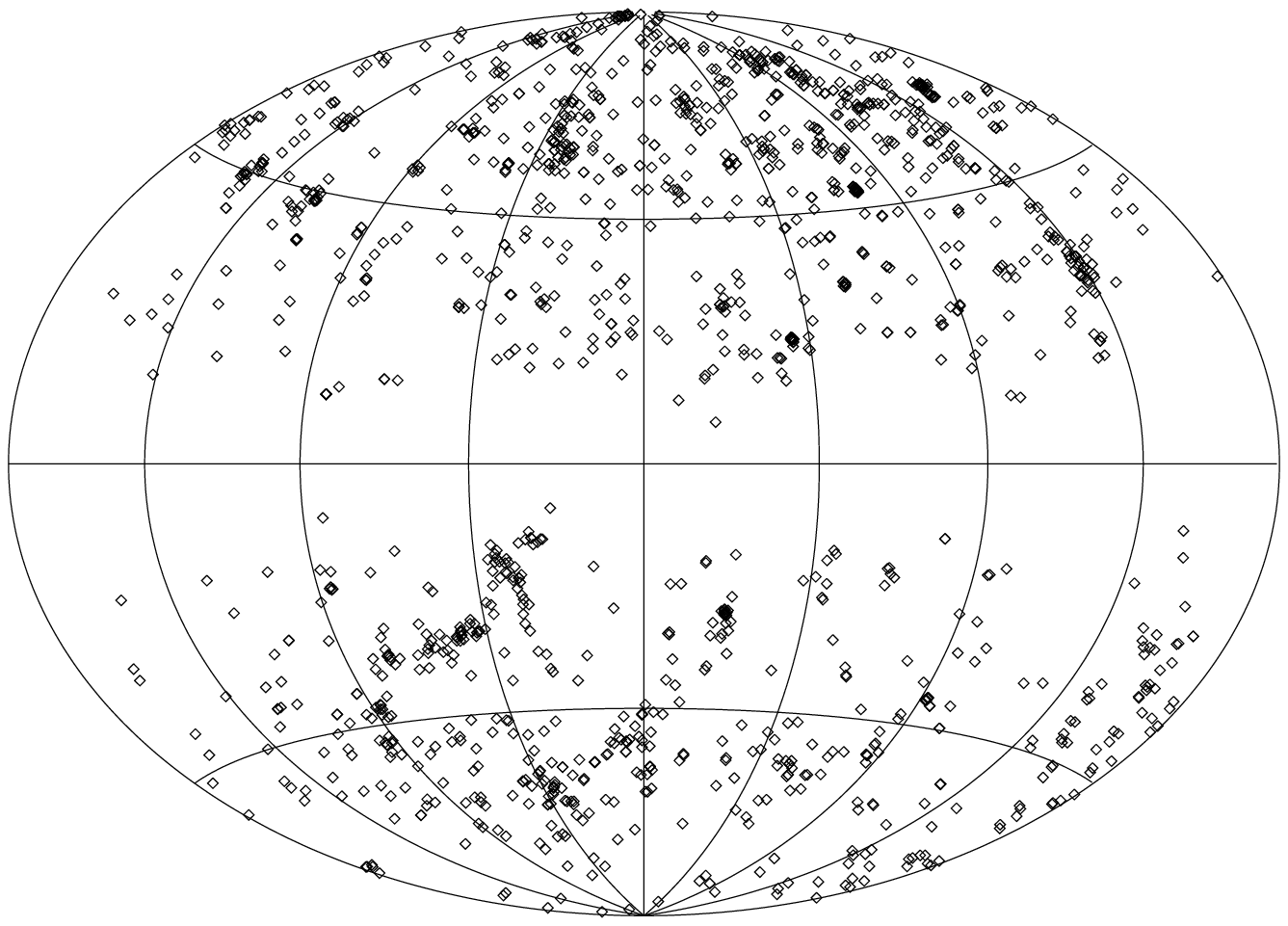}}
\caption{Distribution of SNe in the sky in equatorial (top) and galactic
(bottom) coordinates.}\label{sky}
\end{figure}

In Fig.~\ref{sky} the sky distribution of the 1447 supernovae both in
equatorial (top panel) and galactic (bottom panel) coordinates is
shown.  Neglecting the avoidance zone defined by the galactic plane,
the outcome of the SN searches of the last years makes the SN sky
distributions more homogeneous compared with the same plot reported in
ASC89. Overposed to the clustered pattern of the distribution of
nearby parent galaxies is evident a smooth background component due to
the high-z SNe.

\begin{figure}
\resizebox{\hsize}{!}{\includegraphics*[angle=-90.]{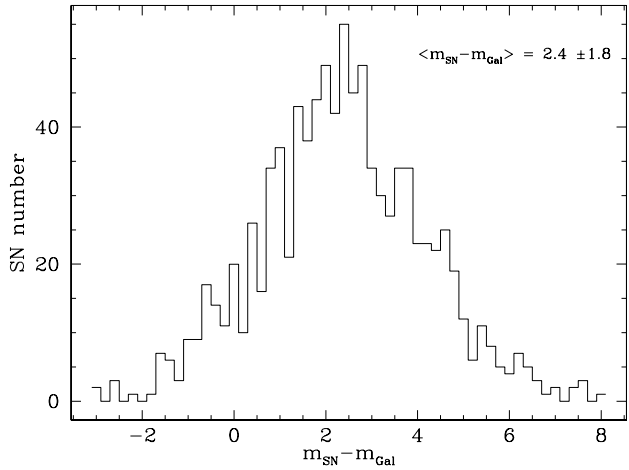}}
\caption{Histogram of the difference between the SN and galaxy magnitudes}
\label{sngal}
\end{figure}

Fig.~\ref{sngal} shows the distribution of the difference between the
SN and parent galaxy magnitude. The peak of the distribution is at
(m$_{SN}$-m$_{Gal})$=2.4 with $\sigma=1.8$ which fairly compares with
the same result found by Barbon (1968). These numbers may be useful to
prepare the strategy of a SN search in given galaxy samples.

Table~5 shows the distribution of supernovae of different types
according to the morphological type of their parent galaxies. With
respect to Table III of ASC89, new SN types are now listed but the
overall distribution remains unchanged.  Note that the percentage of
classified SNe has increased from 40\% to almost 60\% of total
discoveries.

\begin{figure}
\resizebox{\hsize}{!}{\includegraphics*{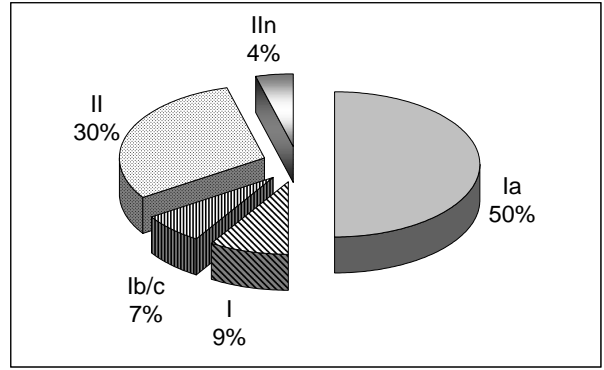}}
\resizebox{\hsize}{!}{\includegraphics*[angle=-90.]{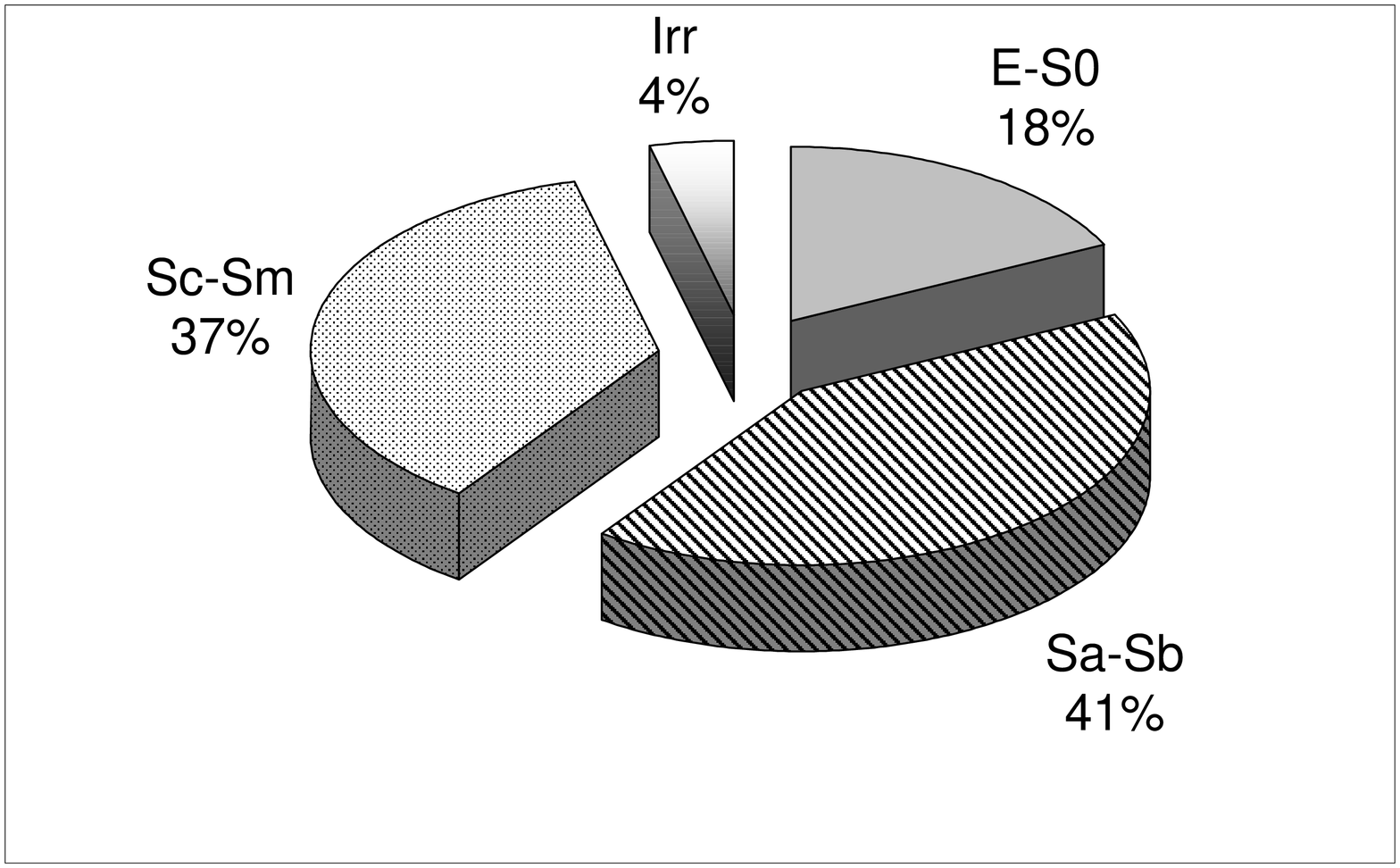}}
\caption{Distributions of SN types (top) and of the parent galaxy morphological types (bottom). }\label{sntype}
\end{figure}

Concerning the distribution among different SN types
(Fig.~\ref{sntype} top), it turns out that type Ia alone make 50\% of
all classified SNe, whereas Ib/c are only 7\%. The same data of ASC89
give 22\% and 6\%. Actually, in ASC89 a major fraction of type I
(54\%) were missing a detailed subtype classification whereas in the
present version this is only 14\%. On the other side, the percentage of
type II SNe (34\%) remain constant.  These numbers show that, since
the last decade, the chase for the SNIa is well under way.

\begin{table*}
\caption{Distribution of supernovae according to the morphological
types of their parent galaxies}
\begin{center}
\begin{tabular}{lrrrrrrrrrrrrrrrrrrr}
\hline
       &E  &S0 &S0/a &Sa &Sab &Sb &Sbc &Sc &Scd &Sd &Sdm &Sm &S  &I0 &Im &I  &Pec &nc  &Total\\
	   						                          
I      &10 &2  &5    &9  &3   &8  &6   &13 &    &2  &    &   &1  &   &   &   &    &18  & 77\\
Ia     &24 &31 &6    &20 &13  &28 &32  &35 &10  &5  &    &1  &18 &   &   &   &    &187 &410\\ 
Iapec  &3  &4  &     &   &2   &1  &3   &2  &    &1  &    &   &1  &   &   &   &1   &1   &19\\
Ib     &   &   &1    &   &1   &1  &1   &9  &1   &   &    &   &   &   &1  &1  &    &1   &17\\
Ib/c   &   &   &     &   &    &   &2   &2  &    &1  &    &1  &1  &   &   &   &    &2   &9\\
Ic     &   &   &1    &1  &1   &2  &3   &9  &2   &   &    &   &4  &   &   &   &    &4   &27\\
Iac    &   &   &     &   &    &2  &    &   &    &   &    &   &   &   &   &   &    &    &2\\
II     &   &   &2    &7  &5   &32 &23  &73 &12  &8  &3   &1  &11 &1  &2  &6  &    &56  &242\\
IIb    &   &   &     &   &1   &   &    &2  &1   &   &    &1  &   &   &   &   &    &    &5\\
IIn    &   &   &1    &1  &    &5  &4   &10 &1   &   &    &   &1  &   &   &2  &    &9   &34\\
IIpec  &   &   &     &1  &    &1  &    &4  &    &   &    &   &1  &   &   &   &    &2   &9\\
Pec    &   &   &     &1  &1   &1  &    &1  &1   &   &    &   &   &   &   &   &    &2   &7\\
nc     &38 &14 &15   &32 &16  &62 &42  &91 &7   &8  &1   &6  &58 &   &2  &19 &1   &177 &589\\
	   						                          
Total  &75 &51 &31   &72 &43  &143&116 &251&35  &25 &4   &10 &96 &1  &5  &28 &2   &459 &1447\\
\hline
\end{tabular}
\end{center}

Notes: nc means not classified supernovae and/or galaxies
\end{table*}

The distribution of the parent galaxy morphological types is also
shown in Fig.~\ref{sntype} (bottom). It stands out clearly that most
SNe are found in spiral galaxies.

\begin{figure}
\resizebox{\hsize}{!}{\includegraphics*[angle=-90.]{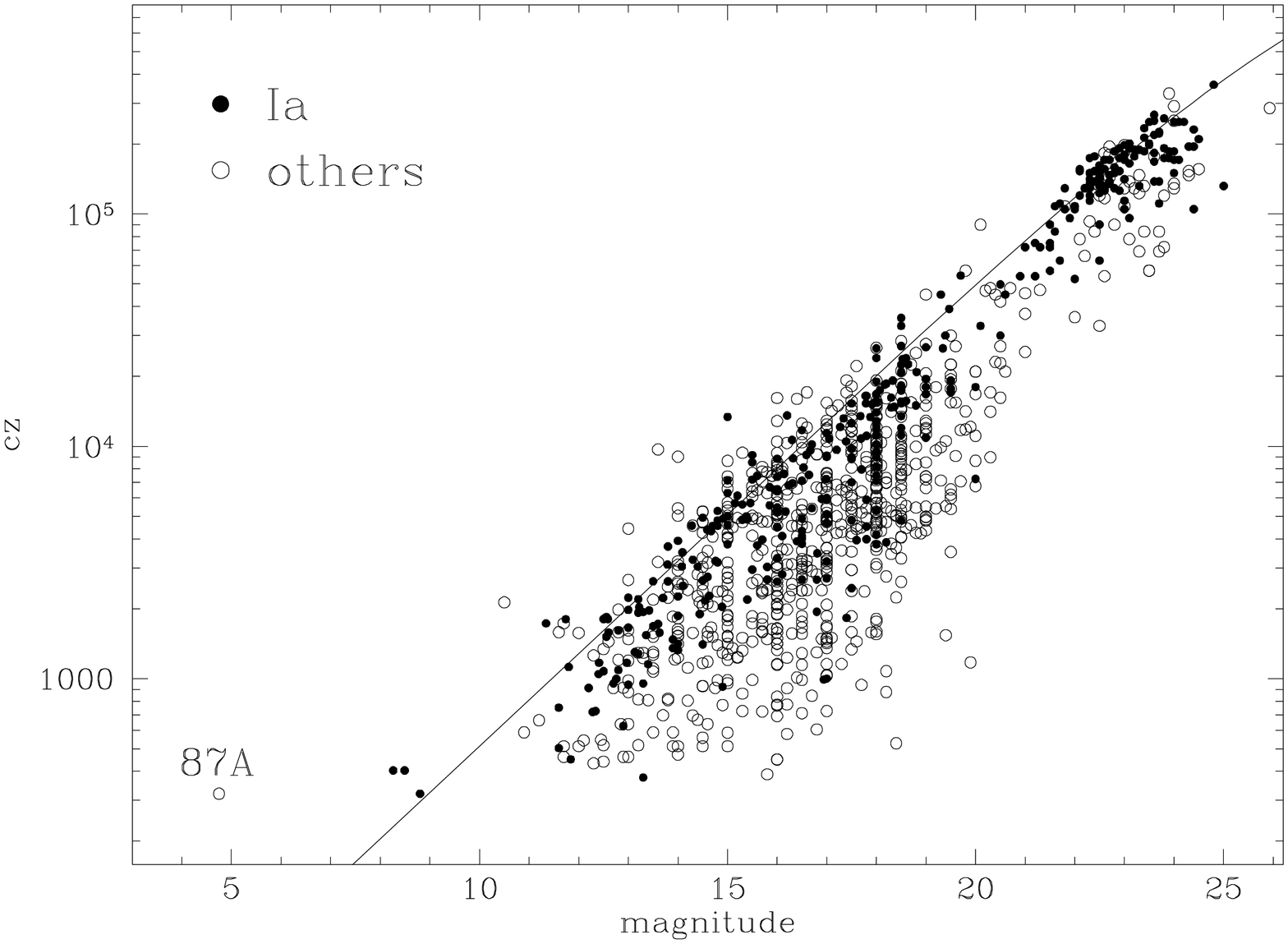}}
\caption{The Hubble diagram for all the SNe with redshift discovered up to Dec. 31, 1998.
The magnitudes are those reported in the Catalogue, i.e. those at
maximum when available, or those at discovery. The line is the
expected position for "standard" SNIa having M(max)=-19.50, Ho=65
km/s~Mpc, qo=0} \label{hubble}
\end{figure}

Finally, in Fig.~\ref{hubble} we plot the positions of all SNe with
redshift in the Hubble diagram. The line is the expected location for
"standard" SNIa having M(max)$=-19.50$, Ho=65 km/s~Mpc, qo=0.  We
remark that some SNe, not of type Ia, laying above
this line have poor photometry.

\begin{acknowledgements}
We are indebted to David Branch for having provided a number of corrections
to a previous release of the Catalogue.
\end{acknowledgements}


\end{document}